\documentclass[prc,showpacs]{revtex4}
\usepackage{graphicx}
\usepackage{dcolumn}
\usepackage{bm}

\begin{document}

\title{Shape mixing and beta-decay properties of neutron-deficient Kr and Sr isotopes}

\author{P. Sarriguren}
\affiliation{Instituto de Estructura de la Materia, CSIC, Serrano
123, E-28006 Madrid, Spain}


\date{\today}

\begin{abstract} 

Gamow-Teller strength distributions and $\beta$-decay half-lives
in neutron-deficient Kr and Sr isotopes are investigated within a 
deformed quasiparticle random phase approximation. The approach
is based on a selfconsistent Skyrme Hartree-Fock mean field with 
pairing correlations and residual separable particle-hole and 
particle-particle forces. A simple two-level model is used to mix 
the nuclear shapes into the physical ground state. Good agreement 
with experiment is found with shape mixing coefficients which are 
consistent with those obtained phenomenologically from mixing of
rotational bands.

\end{abstract}

\pacs{ 21.60.Jz, 23.40.Hc,  27.50.+e}

\maketitle

\section{Introduction}

Neutron-deficient isotopes in the $A=70-80$ mass region are known to
be interesting examples where the equilibrium shape of the nucleus is
the result of a critical interplay between various nuclear structure
effects \cite{wood92}. The evolution of the nuclear shape in isotopic
chains is rather involved due to the existence of shell gaps at nucleon
numbers 34, 36, 38, and 40. Thus, adding or removing a few nucleons
may lead to shape transitions in neighboring nuclei. 
Shape coexistence is also present in this mass region, where competing
prolate, oblate, and spherical shapes are expected in the same nucleus
at close energies. Calculations of the equilibrium configurations in
this mass region have been performed within different approaches, such
as the configuration-dependent shell-correction approach with deformed
Woods-Saxon potentials \cite{naza85}, selfconsistent deformed Skyrme
mean-field calculations  \cite{bonche85,bender}, Hartree-Fock-Bogoliubov
calculations with the Gogny force \cite{hilaire}, relativistic mean-field
calculations \cite{lala95}, or the selfconsistent complex excited VAMPIR
approach \cite{petro_coex}.

The first experimental evidence of shape coexistence in neutron-deficient
Kr isotopes was reported in Ref. \cite{piercey} from irregularities
observed at the bottom of the rotational bands. Since then, a great deal
of data on the low-lying excitation spectrum, including transition
probabilities, have become available \cite{chandler,becker,bouchez,clement}.
One of the experimental indications revealing the existence of shape
coexistence in even-even nuclei is the observation of low-lying $0^+$
excited states. Each of the $0^+$ states is interpreted as a ground
state corresponding to a different shape.
Experimental evidences for $0^+$ shape isomers in Kr isotopes have
been reported in Refs. \cite{chandler,becker,bouchez}. These states
are connected by electric monopole ($E0$) transitions, whose strength
is related to the change in the rms radius of the nucleus between
initial and final states and carries information about the change in
deformation and the overlap of the wave functions. The relationship
between $E0$ strength and shape mixing has been investigated in 
Refs. \cite{heyde,mach,wood99}. In these references, shape mixing and
shape coexistence were analyzed in terms of a two-level model, which
was shown to be a simple but successful method to interpret the
phenomenology. In this model, the strength is related to the amount of
mixing of configurations with different deformations in the physical
states. This simple two-level mixing model has been also successfully
applied to understand the low-energy spectra of neutron-deficient Kr
isotopes \cite{clement}. In these isotopes, regular bands are observed
at high spins, which is a characteristic of well deformed shapes, but
for the lowest states the regularity is lost, which is interpreted as
evidence for a perturbation from other close states. 
The anomaly at low spin observed in the systematics of the moments 
of inertia (see Fig. 3.37 in Ref. \cite{wood99}) has also been interpreted 
as an evidence for mixed ground states in the lightest Sr isotopes. 
This mixing predicts that $E0$ transition strengths should be observable in 
neutron-deficient Sr isotopes. An experiment similar to the study 
of the $^{74}$Rb decay to $^{74}$Kr \cite{pie03}
would be very helpful for the $^{78}$Y decay to $^{78}$Sr.

The decay properties of nuclei in this mass region have been also
investigated both theoretically 
\cite{hamamoto,sarr1,petro_bgt_72,petro_bgt_74} and experimentally 
\cite{piqueras,poirier,nacher,rubio}. Deformation has been identified
in those works as a relevant issue to understand $\beta$-decay
properties. In particular, the nuclear deformation is crucial to
perform reliable calculations of nuclear reaction rates and 
$\beta$-decay half-lives in nuclei involved in the rapid-proton
capture (rp) process of relevance in X-ray burst 
scenarios \cite{schatz,sarri_wp}.

In this work we investigate the decay properties of neutron-deficient
Kr ($^{72,74}$Kr) and Sr ($^{76,78}$Sr) isotopes within a deformed
Hartree-Fock (HF) with Skyrme interactions and pairing correlations
in BCS approximation. Residual spin-isospin interactions are also
included and treated in quasiparticle random phase approximation 
(QRPA) \cite{sarr1,moller_krum,homma,hir}. In Sec. II a brief review
of the theoretical formalism is presented. Sec. III contains the
results obtained within this approach for the potential energy curves,
electric monopole strengths, Gamow-Teller (GT) strength distributions,
and $\beta$-decay half-lives. The results are discussed in terms of
shape mixing using a two-level model. Sec. IV summarizes the main
conclusions.

\section{Theoretical Formalism}

In this section we summarize briefly the theory involved in the
microscopic calculations. More details can be found in 
Ref. \cite{sarr1}. The method consists in a selfconsistent formalism
based on a deformed Hartree-Fock mean field obtained with Skyrme
interactions, including pairing correlations in the BCS approximation.
The single-particle energies, wave functions, and occupation
probabilities are generated from this mean field. Two Skyrme forces
are considered in this paper to quantify the theoretical uncertainties
caused by the use of different effective interactions. One is the
force Sk3 \cite{sk3}, which is one of the simplest and oldest 
parametrizations, and has been successfully tested against many 
nuclear properties in spherical and deformed nuclei. The other force
is SLy4 \cite{sly4}, which is an example of one of the most recent
parametrizations including selected properties of unstable nuclei in
the adjusting procedure.

The solution of the HF equation is found using the formalism developed
in Ref. \cite{vautherin}, assuming time reversal and axial symmetry.
The single-particle wave functions are expanded in terms of the
eigenstates of an axially symmetric harmonic oscillator in cylindrical
coordinates, using twelve major shells. The method also includes
pairing between like nucleons in BCS approximation with fixed gap 
parameters for protons and neutrons, which are determined
phenomenologically from the odd-even mass differences through a 
symmetric five term formula involving the experimental binding
energies \cite{audi}. 

The energy surfaces are analyzed as a function of the quadrupole
deformation. For that purpose, constrained HF calculations are
performed with a quadratic constraint \cite{constraint}. The HF
energy is minimized under the constraint of keeping fixed the 
nuclear deformation. Calculations for GT strengths are performed 
subsequently for the equilibrium shapes of each nucleus, that is, 
for the solutions, in general deformed, for which minima are
obtained in the energy surfaces. Since decays connecting different
shapes are disfavored, similar shapes are assumed for the ground
state of the parent nucleus and for all populated states in the
daughter nucleus. 

To describe GT transitions, a spin-isospin residual interaction
is added to the mean field. This interaction contains two parts,
particle-hole ($ph$) and particle-particle ($pp$). The interaction in
the $ph$ channel is responsible for the position and structure of
the GT resonance \cite{sarr1,homma} and it can be derived consistently
from the same Skyrme interaction used to generate the mean field,
through the second derivatives of the energy density functional with
respect to the one-body densities. The $ph$ residual interaction is
finally written in a separable form by averaging the Landau-Migdal
resulting force over the nuclear volume, as explained in 
Ref. \cite{sarr1}. The $pp$ part is  a neutron-proton pairing force
in the $J^\pi=1^+$ coupling channel, which is also introduced as a 
separable force \cite{hir}.

\begin{equation}
V^{ph}_{GT} = 2\chi ^{ph}_{GT} \sum_{K=0,\pm 1} (-1)^K \beta ^+_K 
\beta ^-_{-K}, \qquad 
\beta ^+_K = \sum_{\pi\nu } \left\langle \nu | \sigma _K |
\pi \right\rangle a^+_\nu a_\pi \, ,
\end{equation}

\begin{equation}
V^{pp}_{GT} = -2\kappa ^{pp}_{GT} \sum_K (-1)^K P ^+_K P_{-K}, \qquad 
P ^+_K = \sum_{\pi\nu} \left\langle \pi \left| \left( \sigma_K
\right)^+ \right|\nu \right\rangle  a^+_\nu a^+_{\bar{\pi}} \, .
\end{equation}
The coupling strengths used in this work are $\chi ^{ph}_{GT}=0.17$
MeV and $\kappa ^{pp}_{GT}=0.03$ MeV.

The proton-neutron QRPA phonon operator for GT excitations in
even-even nuclei is written as

\begin{equation}
\Gamma _{\omega _{K}}^{+}=\sum_{\pi\nu}\left[ X_{\pi\nu}^{\omega _{K}}
\alpha _{\nu}^{+}\alpha _{\bar{\pi}}^{+}+Y_{\pi\nu}^{\omega _{K}}
\alpha _{\bar{\nu}} \alpha _{\pi}\right]\, ,  \label{phon}
\end{equation}
where $\alpha ^{+}\left( \alpha \right) $ are quasiparticle creation
(annihilation) operators, $\omega _{K}$ are the QRPA excitation energies, 
and $X_{\pi\nu}^{\omega _{K}},Y_{\pi\nu}^{\omega _{K}}$ the forward and 
backward amplitudes, respectively. For even-even nuclei the allowed GT
transition amplitudes in the intrinsic frame connecting the QRPA
ground state 
$\left| 0\right\rangle \ \ \left[ \Gamma _{\omega _{K}} \left| 0
\right\rangle =0 \right]$ to one-phonon states $\left| \omega _K 
\right\rangle \ \ \left[ \Gamma ^+ _{\omega _{K}} \left| 0
\right\rangle = \left| \omega _K \right\rangle \right]$, 
are given by

\begin{equation}
\left\langle \omega _K | \sigma _K t^{\pm} | 0 \right\rangle = 
\mp M^{\omega _K}_\pm \, ,
\end{equation}
where
\begin{equation}
M_{-}^{\omega _{K}}=\sum_{\pi\nu}\left( q_{\pi\nu}X_{\pi
\nu}^{\omega _{K}}+ \tilde{q}_{\pi\nu}Y_{\pi\nu}^{\omega _{K}}
\right) , \qquad  M_{+}^{\omega _{K}}=\sum_{\pi\nu}\left( 
\tilde{q}_{\pi\nu} X_{\pi\nu}^{\omega _{K}}+
q_{\pi\nu}Y_{\pi\nu}^{\omega _{K}}\right) \, ,
\end{equation}
with
\begin{equation}
\tilde{q}_{\pi\nu}=u_{\nu}v_{\pi}\Sigma _{K}^{\nu\pi },\ \ \ 
q_{\pi\nu}=v_{\nu}u_{\pi}\Sigma _{K}^{\nu\pi},
\label{qs}
\end{equation}
$v'$s are occupation amplitudes ($u^2=1-v^2$) and 
$\Sigma _{K}^{\nu\pi}$ spin matrix elements connecting neutron 
and proton states with spin operators
\begin{equation}
\Sigma _{K}^{\nu\pi}=\left\langle \nu\left| \sigma _{K}\right| 
\pi\right\rangle \, .
\end{equation}

The solutions of the QRPA equations are found by solving first a
dispersion relation of fourth order in the excitation energies 
$\omega$. The technical procedure to solve the QRPA equations is
described in detail in Ref. \cite{hir}. The Ikeda sum rule \cite{ikeda}
is always fulfilled in these calculations.

The GT strength $B_{\omega}(GT^\pm)$ in the laboratory system for
a transition  $I_iK_i (0^+0) \rightarrow I_fK_f (1^+K)$ can be
obtained as
\begin{equation}
B_{\omega}(GT^\pm ) = \sum_{\omega_{K}} \left[ \left\langle \omega_{K=0} 
\left| \sigma_0t^\pm \right| 0 \right\rangle ^2 \delta (\omega_{K=0}-
\omega ) + 2 \left\langle \omega_{K=1} \left| \sigma_1t^\pm \right| 
0 \right\rangle ^2 \delta (\omega_{K=1}-\omega ) \right] \, ,
\label{bgt}
\end{equation}
in $[g_A^2/4\pi]$ units. To obtain this expression, the initial and
final states in the laboratory frame have been expressed in terms of
the intrinsic states using the Bohr-Mottelson factorization \cite{bm}.
The effect of angular momentum projection is  then, to a large extent, 
taken into account. 

The $\beta$-decay half-life is obtained by summing up all the allowed 
transition probabilities weighted with some phase space factors up to 
states in the daughter nucleus with excitation energies lying below 
the corresponding $Q$-energy,   

\begin{equation}
T_{1/2}^{-1}=\frac{1}{D}\sum_{\omega }f\left( Z,\omega \right)
\left\{ B_{\omega}(F)+ \left[ 0.77\left( g_{A}/g_{V}\right) _{\rm free}
\right] ^{2} B_{\omega}(GT) \right\} \, ,
 \label{t12}
\end{equation}
where $D=6200$~s and 0.77 is a standard quenching factor that takes into
account in an effective way all the correlations \cite{hama_eff} which 
are not properly considered in the present approach. Since Fermi 
contributions are very small, the bare results can be recovered by 
scaling the results in this paper for $B(GT)$ and $T_{1/2}$ with the 
square of this quenching factor. 

In $\beta^+/EC$ 
decay, the Fermi integral $f(Z,\omega)$ consists of two parts, positron
emission and electron capture. In this work they are computed numerically
for each value of the energy, as explained in Ref. \cite{gove}.
The inclusion of the Fermi strength 
$ B_{\omega}(F^+) = [g_V^2/4\pi] < \omega | t^+ | 0 > ^2 $ in the 
$\beta^+/EC$ half-lives becomes important only for nuclei with $Z>N$.
The calculation of these contributions in our case shows that they are
negligible in the $N=Z+2$ isotopes ($^{74}$Kr,$^{78}$Sr) and represents
only a few percent  correction in the case $N=Z$ ($^{72}$Kr,$^{76}$Sr).

\section{Results}

\subsection{Potential energy curves and electric monopole strength}

\begin{figure*}[t]
\centering
\includegraphics[width=150mm]{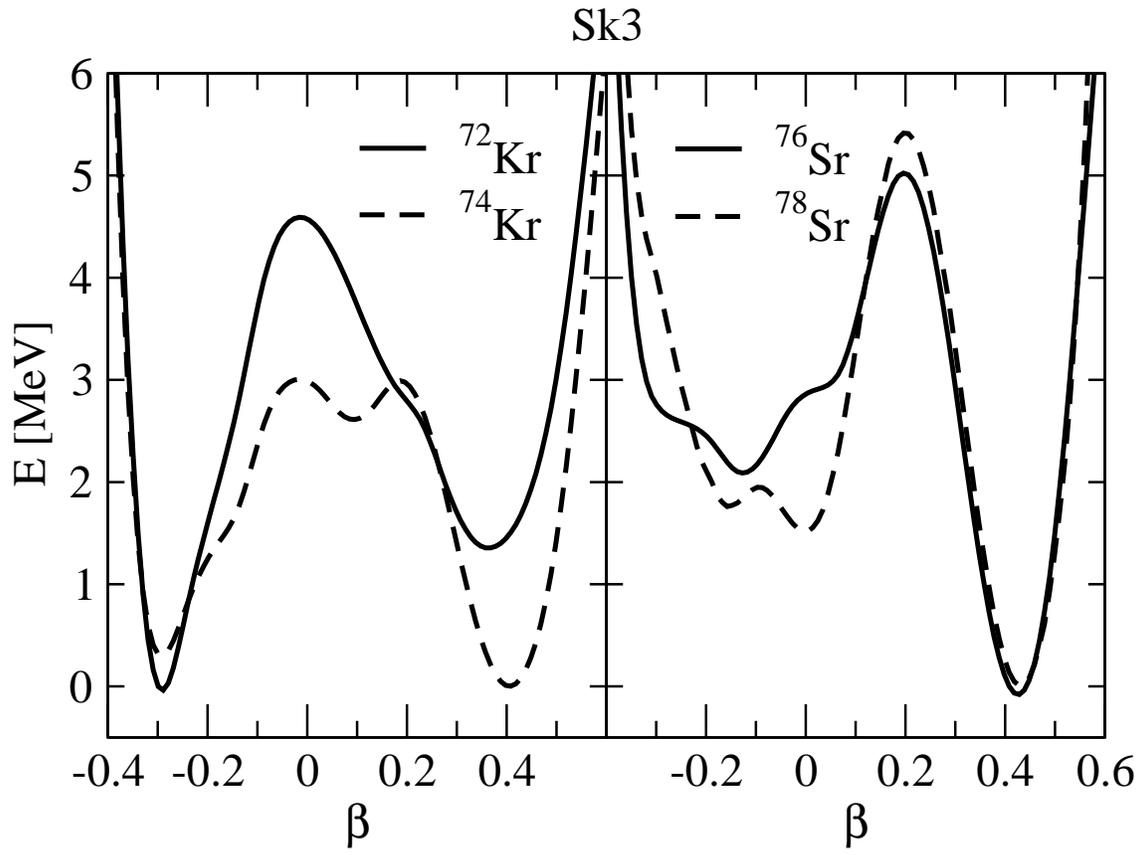}
\caption{Potential energy curves for $^{72,74}$Kr (left panel) and 
$^{76,78}$Sr (right panel), obtained from constrained HF+BCS calculations
with the Skyrme force Sk3.  Beta is the quadrupole deformation.}
\label{fig_eq_sk3}
\end{figure*}

\begin{figure*}[t]
\centering
\includegraphics[width=150mm]{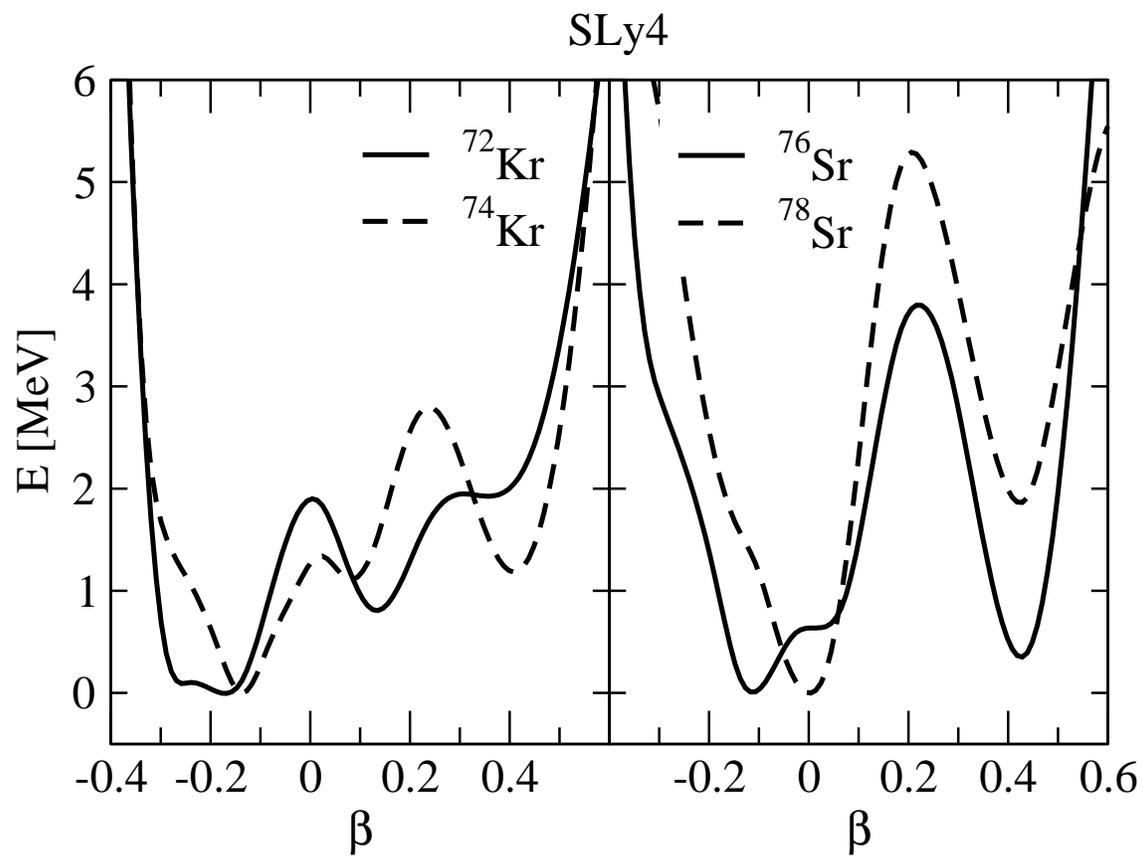}
\caption{Same as in Fig. \ref{fig_eq_sk3} for SLy4.}
\label{fig_eq_sly4}
\end{figure*}

The potential energy curves corresponding to the force Sk3 (SLy4)
for the isotopes $^{72,74}$Kr and $^{76,78}$Sr can be seen in 
Fig. \ref{fig_eq_sk3} (\ref{fig_eq_sly4}). We obtain an oblate ground
state in $^{72}$Kr with a prolate local minimum at higher energy. In
the case of  $^{74}$Kr, the oblate and prolate solutions are practically
degenerate in the Sk3 case, while SLy4 predicts an oblate shape in the
ground state. The Sr isotopes show for Sk3 a prolate ground state and a
second oblate (spherical) minimum for  $^{76}$Sr ($^{78}$Sr). In the
case of SLy4 the oblate and prolate solutions are practically degenerate 
in  $^{76}$Sr. The spherical shape becomes the ground state in the case 
of  $^{78}$Sr. These results are in qualitative agreement with similar
results obtained in this mass region from different methods 
\cite{naza85,bonche85,bender,hilaire,lala95,petro_coex,moller_nix}. 
In all of these studies $^{72}$Kr is found to be oblate in its ground
state with a prolate solution at higher energy. $^{74}$Kr is found to
exhibit prolate and oblate solutions close in energy. Both $^{76}$Sr
and $^{78}$Sr show prolate deformations in their ground states with
oblate and spherical minima at higher energies, respectively. These
features are also in agreement with experiment, where similar shape 
changes have been observed in this region \cite{clement,lister}.

As it was mentioned in the Introduction, the two-level mixing model
has been successfully used to interpret the low-lying excitation
spectra of neutron-deficient Kr isotopes measured in Coulomb
excitation experiments \cite{clement}. Coulomb excitation is a
suitable method to distinguish between different shapes of the nucleus
and to verify the shape coexistence scenario in light Kr isotopes. The
analysis carried out in Ref. \cite{clement} was based on the assumption
of two regular rotational bands with collective intra-band transitions
and zero matrix elements between the intrinsic states of the different
bands. It confirms the prolate-oblate shape coexistence scenario in 
$^{74}$Kr, with an extracted squared mixing amplitude for the prolate
configuration $\lambda =0.48$. The squared amplitude is only 
$\lambda =0.10$ (mostly oblate) in  $^{72}$Kr, as extracted 
phenomenologically in Ref. \cite{bouchez}.

The electric monopole operator can be expressed in terms of 
single-nucleon degrees of freedom as

\begin{equation}
T(E0)= \sum_k e_kr^2_k \, .
\end{equation}
The diagonal matrix elements of this operator give information about
nuclear radii. The nondiagonal matrix elements give $E0$ transition
amplitudes. If we consider the mixing of two  $0^+$  intrinsic
deformed configurations, the mixed or physical states (ground state
and excited state) can be written as a linear combination of them,

\begin{eqnarray}
\left| 0^+_{\rm gs} \right\rangle & =& \sqrt{\lambda} \left| 
0^+_{\rm prol} \right\rangle + \sqrt{1- \lambda} \left| 0^+_{\rm obl} 
\right\rangle \, ,\nonumber \\
\left| 0^+_{\rm exc} \right\rangle & =& \sqrt{1-\lambda} \left| 
0^+_{\rm prol} \right\rangle - \sqrt{\lambda} \left| 0^+_{\rm obl} 
\right\rangle \, ,
\label{prolobl}
\end{eqnarray}
then,

\begin{eqnarray}
&& \rho \left( E0,0_{\rm exc}^{+}\rightarrow 0_{\rm gs}^{+}\right) =
\frac{\left\langle 0_{\rm exc}^{+}\left| \hat E0\right| 0_{\rm gs}^{+}
\right\rangle}{eR^2}  = \frac{1}{eR^{2}} \left[ \sqrt{\lambda 
 \left( 1-\lambda \right) } \right.  \nonumber \\
&& 
\times \left.\left( \left\langle
0_{\rm prol}^{+}\left| \hat{E}0\right| 0_{\rm prol}^{+}\right\rangle 
-\left\langle 0_{\rm obl}^{+}\left| \hat{E}0\right| 0_{\rm obl}^{+}
\right\rangle \right) -\left( 2\lambda -1\right) \left\langle 
0_{\rm prol}^{+}\left| \hat{E}0\right| 0_{\rm obl}^{+}\right\rangle 
\right] \, . 
\label{rhoE0} 
\end{eqnarray}
The cross term is neglected because the wave functions are mainly
localized at different points in deformation space \cite{heyde}.
This expression can be written in terms of the square  of the 
difference between the rms radii of the states which are mixed.
Using an expansion of the radii in the deformation parameter $\beta$
up to third order, this expression can also be written 
as \cite{rhobeta,heyde} 

\begin{equation}
\rho ^2 \left( E0,  0_{\rm exc}^{+}\rightarrow 0_{\rm gs}^{+}
\right)= \lambda (1-\lambda) \left( \frac{3Z}{4\pi} \right) ^2
\left[ \left( \beta_{\rm prol} ^2 - \beta_{\rm obl} ^2 \right) 
+\frac{5}{21} \sqrt{\frac{5}{\pi}} \left( \beta_{\rm prol} ^3 
-\beta_{\rm obl} ^3 \right) \right] ^2 \, .
\label{rho_beta}
\end{equation}
As one can see, if there is no mixing, the $E0$ strength becomes zero
and it is largest for maximal mixing. If the deformations  $\beta $
are known, one can extract from these expressions the mixing of the
two deformed configurations. Dynamical effects beyond mean-field
approach have been considered in various works \cite{bender,dancer}.
The results obtained from the Generator Coordinate Method and those
from the phenomenological two-level model are compared in the mass
region $A=190$. The monopole strengths calculated with both methods
agree quite well \cite{dancer}.

\begin{figure*}[t]
\centering
\includegraphics[width=120mm]{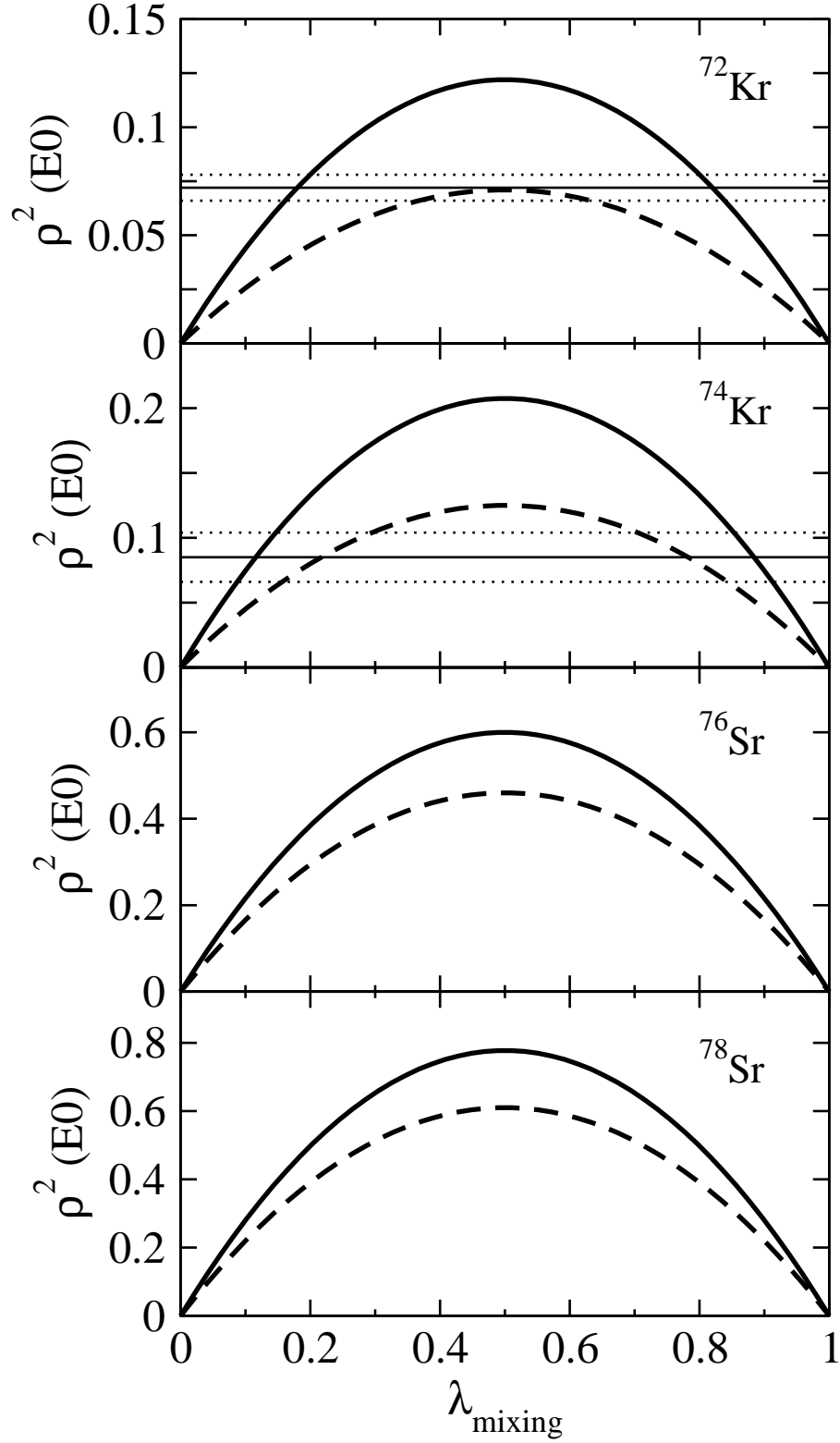}
\caption{Electric monopole strength $\rho^2 (E0)$ with Sk3 as a function
of the mixing parameter $\lambda$, calculated from Eq. (\ref{rho_beta})
with (solid lines) and without (dashed lines) $\beta^3$ terms. Horizontal
lines in $^{72}$Kr and  $^{74}$Kr correspond to the experimental 
values from \cite{bouchez} and \cite{chandler}, respectively. }
\label{fig_rho}
\end{figure*}

Fig. \ref{fig_rho} contains the electric monopole strength 
$\rho^2 (E0)$ obtained with the force Sk3 from Eq. (\ref{rho_beta})
as a function of the mixing parameter $\lambda$. The solid (dashed)
lines correspond to the results obtained with (without) the $\beta^3$
terms in Eq. (\ref{rho_beta}), terms that are neglected in many
works. Horizontal lines in $^{72}$Kr and  $^{74}$Kr are the
experimental values from \cite{bouchez} and \cite{chandler}, 
respectively. One can see that the experimental values in the Kr
isotopes are compatible with  $\lambda=0.1-0.2$, which is consistent
with the mixing extracted in Ref. \cite{bouchez} for $^{72}$Kr 
($\lambda_{\rm exp}=0.1$), but inconsistent with the mixing extracted 
in Ref. \cite{clement} for $^{74}$Kr  ($\lambda_{\rm exp} = 0.48$). 
In the case of Sr isotopes, $\rho^2 (E0)$ strengths are larger than
in the case of Kr isotopes. Assuming that the experimental strengths
are similar to those for Kr isotopes, this would indicate very weak
mixing, which is consistent with the strong prolate component
expected in these isotopes.

\subsection{Energy distributions of the Gamow-Teller strength}

In previous works \cite{sarr1,sarri_wp,sarr2} we have studied the
sensitivity of the GT strength distributions to the various
ingredients contributing to the QRPA-like calculations, namely to 
the NN effective force, to pairing correlations, to deformation, 
and to residual interactions. We found different sensitivities 
to them. In this work, all of these ingredients have been fixed
to the most reasonable choices found previously. Here, we mainly
discuss the mixing of different shapes needed to reproduce the
available experimental information on the energy distributions of
GT strengths. Experimental information on GT strength distributions
is available for $^{72}$Kr \cite{piqueras}, $^{74}$Kr \cite{poirier},
and $^{76}$Sr \cite{nacher}. Measurements on $^{78}$Sr are being
presently analyzed \cite{rubio}. It should be mentioned that data
for $^{74}$Kr, $^{76}$Sr, and $^{78}$Sr were taken using the total
absorption gamma spectroscopy (TAgS) technique that avoids 
systematic uncertainties related to the so-called Pandemonium effect
associated with the high resolution techniques, used for example in
the case of $^{72}$Kr \cite{piqueras}. Thus, while the GT strength
has been observed over most of the $Q_{EC}$ window in the former
cases, in the latter case, only the GT strength below 2 MeV was
extracted, which is still far from the  $Q_{EC}=5.04$ MeV energy.

\begin{figure*}[t]
\centering
\includegraphics[width=150mm]{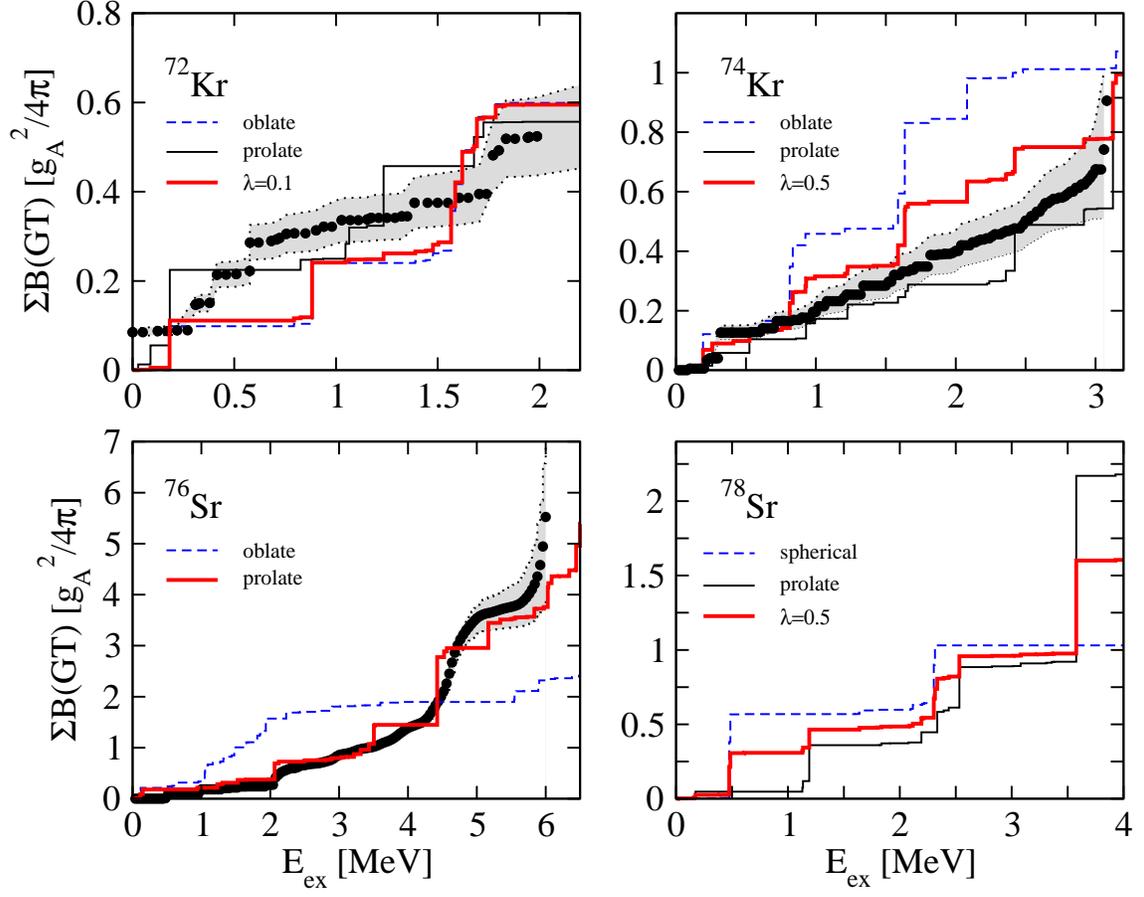}
\caption{(color online) Cumulative QRPA GT strength distributions with
the force Sk3 as a function of the excitation energy in the daughter
nucleus. The calculations correspond to the various equilibrium
configurations and to the adopted mixing. Experimental data (black dots)
are from \cite{piqueras} ($^{72}$Kr), \cite{poirier} ($^{74}$Kr), and
\cite{nacher} ($^{76}$Sr). }
\label{fig_bgt_sk3}
\end{figure*}

\begin{figure*}[t]
\centering
\includegraphics[width=150mm]{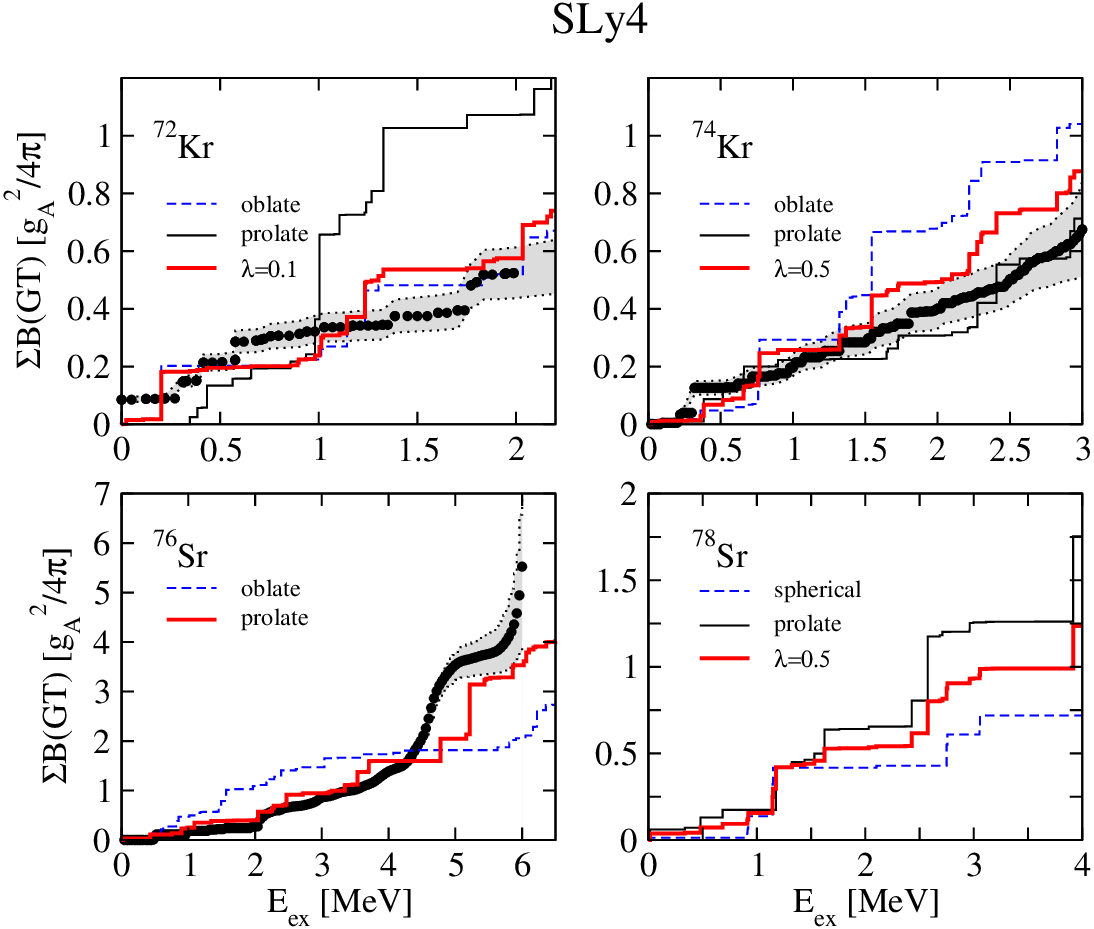}
\caption{(color online) Same as in Fig. \ref{fig_bgt_sk3}, but using SLy4 force.}
\label{fig_bgt_sly4}
\end{figure*}

Figures \ref{fig_bgt_sk3} and \ref{fig_bgt_sly4} show the 
cumulative GT strength distributions as a function of the excitation
energy in the daughter nucleus. Data are compared with Sk3 and SLy4
calculations corresponding to the various equilibrium configurations
and to the adopted mixing. A quenching factor 0.77 is included in
these calculations. 
When the ground state of the parent nucleus is given by a superposition
of prolate and oblate shapes (Eq.(\ref{prolobl})), the final $1^+$ 
states reached by the action of the GT operator are either $|1^+_{\rm prol}>$
or  $|1^+_{\rm obl}>$ at excitation energies given by $\omega_{\rm prol}$
and $\omega_{\rm obl}$, respectively. Thus,

\begin{equation}
B_{\omega}(GT) = \lambda \sum_{\omega_{\rm prol}} \left\langle 
1^+_{\rm prol},\omega_{\rm prol} \left| GT \right| 0^+_{\rm prol} 
\right\rangle ^2 \delta (\omega_{\rm prol}-\omega ) + (1-\lambda)
\sum_{\omega_{\rm obl}} \left\langle 
1^+_{\rm obl},\omega_{\rm obl} \left| GT \right| 0^+_{\rm obl} 
\right\rangle ^2 \delta (\omega_{\rm obl}-\omega ) \, .
\end{equation}
This is certainly a simple approach that gives us only a limited
insight into the more involved problem of configuration mixing,
but it is a first step in this direction toward the effect of mixing
in the Gamow-Teller strength distributions.

The mixing coefficients considered in these plots
correspond to the experimental mixing obtained in 
Refs. \cite{clement,bouchez} for the Kr isotopes. In the case of 
$^{76}$Sr no mixing is plotted since the prolate configuration
reproduces the data fairly well. In the case of $^{78}$Sr, 50\% 
mixing is plotted as a middle value useful to compare with future
data.
In the case of  $^{72}$Kr, the description of the GT strength 
distribution  in the case of Sk3 is rather similar to both shapes.
On the contrary, with SLy4, the oblate shape and in particular the
mixing with a 10\% prolate configuration describes the data fairly. 
Certainly, it would be interesting to extend the measurements
up to the $Q_{EC}$ window and compare the calculations in the whole
range of energies. In the case of $^{74}$Kr, the experiment is better
reproduced by  the prolate shape with the two Skyrme forces. However,
mixing with the oblate shape improves the results although 50\% mixing
seems to be very strong.

The prolate configuration alone gives a good description of the GT
strength distributions  in $^{76}$Sr with both Sk3 and SLy4 forces. 
The oblate shape generates a rather flat profile above 2 MeV that
fails to account for the experimental profile of the strength
distribution. The calculated strength distributions in $^{78}$Sr show
a very pronounced stepwise profile in the spherical case, as it
corresponds to transitions between degenerate states. In the prolate
case the strength is more fragmented. The total accumulated strength 
up to about 3 MeV is similar with the two forces, but in the case of
Sk3, there is a strong transition just below $Q_{EC}$ which doubles
the strength. In the case of SLy4, this strong transition occurs a
little bit above $Q_{EC}=3.76$ MeV.

These results also can be compared with other calculations such
as those performed within a Tamm-Dancoff approximation with Sk3
interaction \cite{hamamoto}. In this reference, strengths contained 
in bins of 1 MeV were plotted and no further details within the 
$Q_{EC}$ window were shown. Nevertheless, there is qualitative
agreement with our results. More recently, results from the complex
excited VAMPIR variational approach with Bonn potentials, on
$^{72}$Kr and $^{74}$Kr, have been published 
\cite{petro_bgt_72,petro_bgt_74}. In the case of  $^{72}$Kr, the
profile of the cumulative GT strength in Ref. \cite{petro_bgt_72}
presents a strong jump at around 1 MeV with a continuous increase 
elsewhere. This is at variance with experiment which does not show
this behavior. For $^{74}$Kr, again a sudden increase of the strength
at 1 MeV is found in \cite{petro_bgt_74} which does not show up in the 
experiment. The total strength is considerably underestimated by the
calculations even though no quenching is considered in that work.

\subsection{Half-lives}

The half-lives calculated according to Eq.(\ref{t12}) can be seen in 
Table I. The results correspond to the different equilibrium 
deformations of the various isotopes, as well as to the mixed
configurations obtained with similar mixing as those considered in
Figs. \ref{fig_bgt_sk3} and \ref{fig_bgt_sly4} for the GT strength
distributions. Experimental $Q_{EC}$ values have been used in these
calculations.

\begin{figure*}[t]
\centering
\includegraphics[width=150mm]{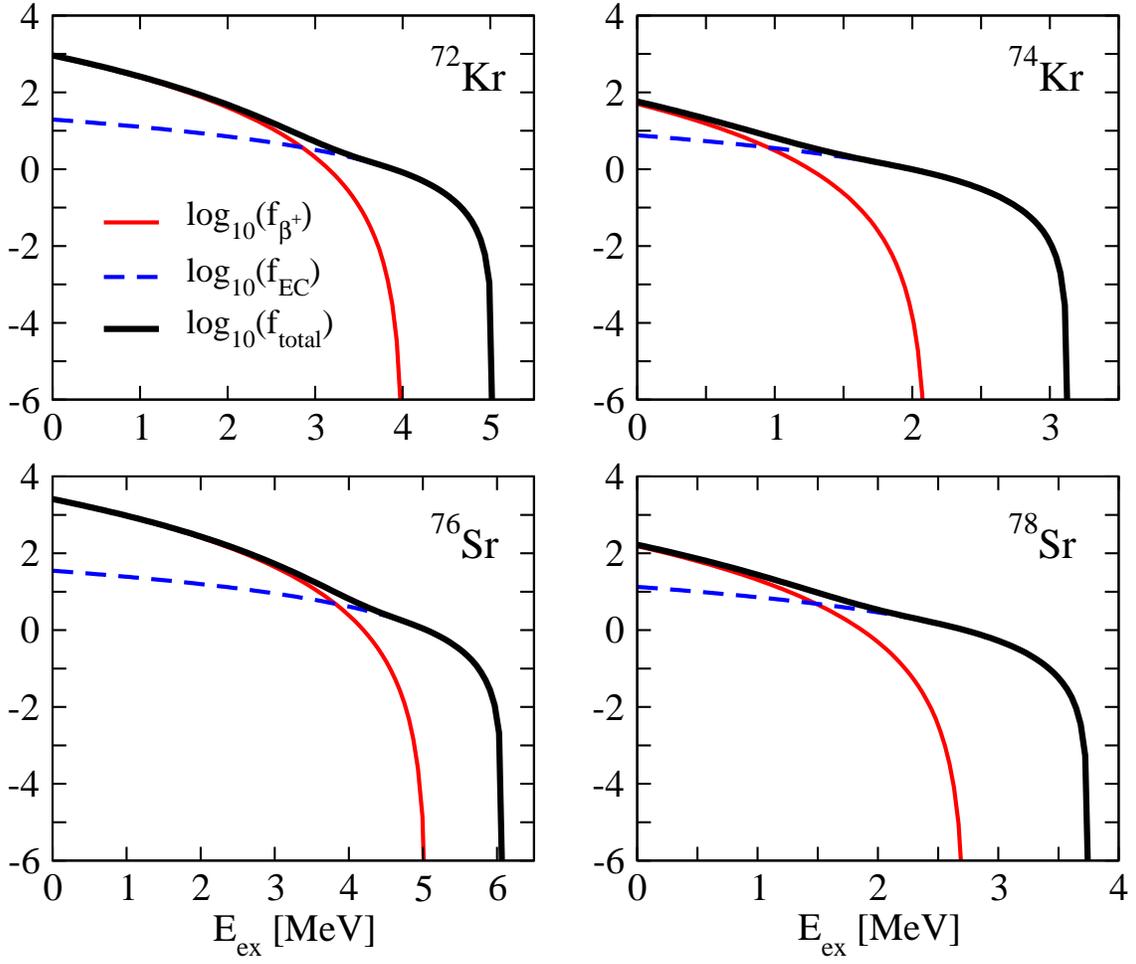}
\caption{(color online) $f(Z,\omega)$ functions used in Eq.(\ref{t12}) to evaluate 
the half-lives. They are decomposed into their $\beta^+$ and electron capture (EC) 
components.}
\label{fig_f_factors}
\end{figure*}

It should be stressed that the $f(Z,\omega)$ functions in
Eq.(\ref{t12}) weight differently the strength depending on the 
energy. As a result, the half-lives are more sensitive to the 
strength located at certain excitation energies. One can see in 
Fig. \ref{fig_f_factors} the $f(Z,\omega)$ functions corresponding
to the $\beta^+$-decay, to the electron capture, and the total 
$f(Z,\omega)$ plotted
versus the excitation energy of the daughter nucleus.
The first thing to notice is that these factors are larger at lower 
excitation energies (higher energy of the $\beta$-particle) and 
therefore, the contribution of the strength at low excitation energies
to the half-lives is highly favored. One can also see that the
$\beta^+$ component is the dominant contribution at lower excitation 
energy and it goes roughly like $(Q_{\beta} - E_{ex})^5$. At higher 
energies, electron capture becomes dominant and it is indeed the only 
component between $Q_{\beta}$ and $Q_{EC}$, where electron capture is 
allowed but positron emission is energetically forbidden.

The results in Table I obtained for $^{72}$Kr are once more compatible 
with predominantly oblate shape with a 10\% prolate mixing \cite{bouchez},
although a little bit more mixing is favored (actually the half-life
is reproduced with $\lambda=0.3$ in the case of Sk3 and with 
$\lambda=0.2$ in the case of SLy4). In the case of $^{74}$Kr one can
see that the 50\% maximal mixing considered also reproduces
the experimental data  reasonably well. In the case of the Sk3 force, 
$\lambda=0.6$
reproduces the half-life. The half-life of $^{76}$Sr is compatible
with a pure prolate shape, as in the case with the GT strength
distribution. In the case of $^{78}$Sr, a mixing of the spherical
and prolate configurations in Sk3 reproduces the half-life, while in
the case of SLy4 a prolate configuration is favored.

\section{Conclusions}

In this paper $\beta$-decay properties of neutron-deficient Kr and Sr
isotopes are calculated within a deformed QRPA approach based on mean
fields generated from selfconsistent Skyrme Hartree-Fock calculations.
A simple two-state mixing model has been used to mix the intrinsic
configurations into the physical states. It is shown that in some cases
a single shape accounts for the main characteristics of the GT
strength distributions and the half-lives. In other cases the data
appear between the predictions of various shapes, demanding a more
sophisticated treatment. Here we have considered a rough estimate
of the mixing using the same coefficients extracted phenomenologically 
from Coulomb excitation experiments and have found that they reproduce 
the GT strength distributions and half-lives fairly well. 
Thus, we get a globally consistent picture 
when describing the isotope $^{72}$Kr as a mainly oblate nucleus with
small admixtures from a prolate configuration, the isotope $^{74}$Kr
as a more mixed nucleus with a dominant prolate configuration, the
isotope $^{76}$Sr as an almost prolate nucleus, and the isotope 
$^{78}$Sr as a mixed spherical/prolate nucleus.

The main objective in this paper has been to demonstrate that this
approach is able to account for the main features of the decay
properties of nuclei in this mass region, which are characterized by
deformation as a key ingredient. This study provides additional and
complementary indications in favor of shape coexistence in this mass
region that is consistent with the information extracted from
low-energy Coulomb excitation experiments.

This approach cannot be pushed forward at a more quantitative
description since the results are not only sensitive to the details 
of the present calculation but may also be sensitive to dynamical 
effects beyond mean-field approach not considered in this work.
In this respect, it is worth pointing out that Multiple 
Coulomb Excitation data are beginning to yield sufficient
numbers of $E2$ matrix elements for construction not only
of quadrupole centroids but also fluctuation widths for
the lowest few states in even-even nuclei \cite{cline}.

\begin{acknowledgments}
This work was supported by Ministerio de Ciencia e Innovaci\'on
(Spain) under Contract No. FIS2008--01301.
\end{acknowledgments}

\newpage

\begin{table}[ht]
\begin{center}
\caption{Half-lives ($T_{1/2}$ [s]) corresponding to the forces  Sk3
and SLy4 for the various shapes and prolate mixing $\lambda$. In the
case of $^{78}$Sr the oblate results correspond actually to the 
spherical configuration.} 
\vskip 0.5cm
{\begin{tabular}{llrrrr} 
\hline
\hline     && oblate & prolate & mixed ($\lambda$ ) & exp. \\
\hline
Sk3 &&&&& \\
& $^{72}$Kr  & 19.7  & 14.0   &  18.9 (0.1)    & 17.2  \\
& $^{74}$Kr  & 441.1 & 1020.0 &  615.9 (0.5)   & 690   \\
& $^{76}$Sr  & 4.2   & 8.9    &  8.9 (1.0)     & 8.9   \\
& $^{78}$Sr  & 93.8  & 320.6  &  145.1 (0.5)   & 150  \\
\\
SLy4 &&&&& \\
& $^{72}$Kr  & 18.2  & 13.7   &  17.6 (0.1)    & 17.2  \\
& $^{74}$Kr  & 718.9 & 858.5  &   782.5 (0.5)  & 690   \\
& $^{76}$Sr  & 5.2   & 10.0   &  10.0 (1.0)    & 8.9   \\
& $^{78}$Sr  & 343.4 & 165.9  &  223.7 (0.5)   & 150  \\
\hline
\end{tabular}}
\end{center}
\end{table}

\newpage

\end{document}